\documentclass[letter]{aa}

\usepackage{amsmath,amssymb}
\usepackage[varg]{txfonts}
\usepackage{graphicx}
\usepackage{natbib}
\bibpunct{(}{)}{;}{a}{}{,}

\usepackage{xspace}
\usepackage[utf8]{inputenc} 
\usepackage{wasysym}
\usepackage{marvosym}
\usepackage{textcomp}
\usepackage{booktabs}

\graphicspath{{figures/}}

\newcommand{\XMM}{\textsl{XMM}\xspace}
\newcommand{\XMMNewton}{\textsl{XMM-Newton}\xspace}

\newcommand{\Fermi}{\textsl{Fermi}\xspace}

\newcommand{\Chandra}{\textsl{Chandra}\xspace}
\newcommand{\Suzaku}{\textsl{Suzaku}\xspace}
\setcitestyle{notesep={ }}

\newcommand{\pmn}{PMN\,J1603$-$4904\xspace}

\newcommand{\ISIS}{\textsc{ISIS}\xspace}

\usepackage{pdflscape}
\makeatletter
\ifcase \@ptsize \relax
  \newcommand{\miniscule}{\@setfontsize\miniscule{7}{8}}
\or
  \newcommand{\miniscule}{\@setfontsize\miniscule{6}{7}}
\or
  \newcommand{\miniscule}{\@setfontsize\miniscule{6}{7}}
\fi
\makeatother

\title{Investigating source confusion in \pmn}
\author{
  F.~Krau\ss{}\inst{\ref{affil:uvaga}}
  \and M.~Kreter\inst{\ref{affil:wuerzburg}}
  \and C.~Müller\inst{\ref{affil:nijmegen}}
  \and A.~Markowitz\inst{\ref{affil:pol},\ref{affil:ucsd}}
  \and M.~B\"ock\inst{\ref{affil:remeis}}
  \and T.~Burnett\inst{\ref{affil:uw}}
  \and T.~Dauser\inst{\ref{affil:remeis}}
  \and M.~Kadler\inst{\ref{affil:wuerzburg}}
  \and A.~Kreikenbohm\inst{\ref{affil:wuerzburg}} 
  \and R.~Ojha\inst{\ref{affil:gsfc}}
  \and J.~Wilms\inst{\ref{affil:remeis}}
}

\institute{
  GRAPPA \& Anton Pannekoek Institute for Astronomy, University of
  Amsterdam,  Science Park 904, 1098 XH Amsterdam, The Netherlands\\
  \email{Felicia.Krauss@uva.nl}
  \label{affil:uvaga}
  \and
  Institut f\"ur Theoretische Physik und Astrophysik, Universit\"at
  W\"urzburg, Emil-Fischer-Str.\ 31, 97074 W\"urzburg, Germany
  \label{affil:wuerzburg}
  \and
  Department ofAstrophysics/IMAPP, Radboud University Nijmegen,
  Heyendaalseweg 135, 6525 AJ Nijmegen, Netherlands
  \label{affil:nijmegen}
  \and
  Nicolaus Copernicus Astronomical Center, Polish Academy of Sciences,
  Bartycka 18, PL-00-716 Warszawa, Poland
    \label{affil:pol}
  \and
   Center for Astrophysics \& Space Science, University of California,
   San Diego, La Jolla, CA, 92093-0424, USA
   \label{affil:ucsd}
      \and
  Dr.~Remeis Sternwarte \& ECAP, Universit\"at Erlangen-N\"urnberg,
  Sternwartstrasse 7, 96049 Bamberg, Germany 
  \label{affil:remeis}
   \and
   Department of Physics, University of Washington, Seattle, WA 98195-1560, USA
   \label{affil:uw}
  \and
  NASA, Goddard Space Flight Center, Astrophysics Science Division,
  Code 661, Greenbelt, MD 20771, USA
  \label{affil:gsfc}  
}

\authorrunning{F. Krau{\ss}~et al.}
\titlerunning{Investigating source confusion in \pmn}
\date{Received 22 November 2017 / Accepted $<$date$>$}

\abstract{
  \pmn is a likely member of the rare class of $\gamma$-ray emitting
  young radio galaxies. Only one other source, PKS\,1718$-$649, has been
  confirmed so far. These objects, which may transition into larger
  radio galaxies, are a stepping stone to understanding AGN evolution.
  It is not completely clear how these young galaxies, seen edge-on, can
  produce high-energy $\gamma$-rays. \pmn has been detected by
  TANAMI Very Long Baseline Interferometry (VLBI) observations and has
  been followed-up with multiwavelength
  observations. A \Fermi/LAT $\gamma$-ray source has been associated
  with it in the LAT catalogs. We have obtained \Chandra observations of
  the source in order to consider the possibility of source confusion,
  due to the relatively large positional uncertainty of \Fermi/LAT.
  The goal was to investigate the possibility of other X-ray bright
  sources in the vicinity of \pmn that could be counterparts to the
  $\gamma$-ray emission. With \Chandra/ACIS, we find no other sources
  in the uncertainty ellipse of \Fermi/LAT data, which includes an
  improved localization analysis of 8 years of data. We further study
  the X-ray fluxes and spectra. We conclude that \pmn is indeed the
  second confirmed $\gamma$-ray bright young radio galaxy.
}


\begin{document}
\maketitle

\section{Introduction}\label{sec-intro}
Active Galactic Nuclei (AGN) are the most luminous persistent objects
in the Universe.
A subset of AGN exhibits relativistic outflows, called jets. Many
questions of jet physics remain unsolved, including the details of jet
launching, confinement and acceleration processes.
In this context, peculiar AGN in transitory stages
become relevant for addressing the key questions of AGN
science.
Among these objects are young radio galaxies, which exhibit shorter
jets (up to a few kpc), and are also known as compact symmetric
objects (CSO) due to their compactness at radio wavelengths
\citep{Phillips1982,Wilkinson1994,Readhead1996,ODea1998}. They are
typically seen at large inclination angles to the jet(s) and have
negligible Doppler boosting.
While $\gamma$-ray emission from young AGN was predicted
\citep{Kino2007,Kino2009,Stawarz2008,Kino2011}, a detection by the
\textsl{Fermi Gamma-ray Space Telescope} Large Area Telescope
(hereafter \Fermi/LAT) remained elusive for many
years \citep{Dammando2016}. Using the improved Pass~8
reconstruction \citep{Pass8}, \cite{Migliori2016} detected the first
CSO at $\gamma$-ray energies, PKS\,1718$-$649. Three other
candidate sources have been proposed: 4C\,$+55.17$
\citep{McConville2011}, PKS\,1413+135 \citep{Gugliucci2005}, and \pmn
\citep{PMN2014}. The first two have not been confirmed to be young
radio galaxies. It remains unclear which attributes make some young
AGN $\gamma$-ray loud. A direct link between Narrow Line Seyfert 1
galaxies (NLS1) and young AGNs (Compact Symmetric Sources) has been
suggested by \citet{Caccianiga2014}, but seems unlikely \citep{Orienti2015}.

PMN\,J1603$-$4904 is a radio source \citep{PMN}. It was recently
confirmed to be a compact symmetric object from MHz data, which makes
it a young or frustated AGN \citep{PMN2016r}. It was detected by
\Fermi/LAT and classified as a low synchrotron peaked (LSP) BL Lac object
\citep[2FGL\,J1603.8$-$4904,
  3FGL\,J1603.9$-$4903;][]{Nolan2012,FGL3,LAC3}. The first association
with the $\gamma$-ray source was proposed by \cite{Kovalev2009}. It is
also listed in the LAT catalogs of sources detected above 10\,GeV and 50\,GeV
\citep[1FHL\,J1603.7$-$4903,2FHL J1603.9-4903, 3FHL
  J1603.8$-$4903][]{Ackermann2013,fhl2,fhl3}. It is reported as
a variable source in the 3FGL catalog.
\pmn was first proposed to be a $\gamma$-ray bright CSO by
\cite{PMN2014}, who discussed its unusual VLBI structure and spectral energy
distribution (SED). We followed up on this paper with X-ray observations with
\Suzaku and \XMMNewton, which led to the first high S/N X-ray spectrum
that also exhibited an emission line at 5.44\,keV, which we interpreted
as a redshifted neutral Fe K$\alpha$ line \citep[at $z=0.18$;][]{PMN2015}.
Optical data by \citet{Shaw2013}, which resulted in the LSP
BL Lac classification, were not sensitive enough to detect any lines.
Further optical spectroscopy showed our proposed redshift to be
incorrect. X-Shooter data resulted in a redshift measurement of
$z=0.2321\pm0.0004$ \citep{PMN2016op}.
The emission line is due to He-like Fe, emitted at a rest frame energy
of 6.7\,keV.
This emission feature is
not typically seen in AGN, where edge-on sources  exhibit neutral or
slightly ionized Fe K$\alpha$ emission, which is expected to
originate in the accretion disks. This feature seems to be common for
CSO sources \citep{Siemiginowska2016}.
Highly ionized Fe emission is also observed in the
LINER galaxy M81 \citep{Page2004}, though the latter still exhibits
neutral Fe K$\alpha$ emission while \pmn does not. It is unclear
whether this suggests a complete lack of an accretion disk, a
truncated accretion disk (which could likely achieve the high
temperatures to ionize Fe) or a lack of neutral Fe.

Further radio studies of \pmn find a low-frequency turnover in the MHz
-- GHz spectrum, which indicates a source extent of 1.4\,kpc and
confirmed its young radio source classification \citep{PMN2016r}.
The source is located close to the Galactic plane ($l=332\fdg15$,
$b=2\fdg57$), which hinders optical/UV and soft X-ray observations
due to extinction and photoelectric absorption. The low Galactic
latitude also complicates $\gamma$-ray data analysis due to the large
number of nearby sources, the Galactic diffuse emission, and the three
nearby extended sources, which all have to be taken into account.
The radio source is consistent with the \Fermi/LAT 95\% 2FGL
positional uncertainty, but it is unclear if there are other possible
counterparts within the uncertainty ellipse in either radio or X-ray
wavelengths \citep{PMN2014}. Optical data are unable to solve this
problem, due to the large number of nearby stars and the strong
extinction.
In this paper we examine recent \Chandra/ACIS data, the
highest angular resolution data available at high energies, to confirm
or rule out source confusion for \pmn. In Section 2 we discuss the
observations and analysis methods. In Section 3 we discuss
the results of the observations. The final section reports our
conclusions.

\section{Observations \& Methods}\label{sec-obs}
\subsection{\Chandra}
We took one \Chandra observation of \pmn with the Advanced CCD Imaging
Spectrometer (ACIS; observation ID~17106, 10.08\,ksec) on 12 May 2016.
The data are not affected by pile-up, as \pmn is a relatively weak
X-ray emitter \citep{PMN2015}.
The data were extracted using the standard tools from the CIAO\,4.8.
The extraction radius for \pmn is 3\farcs3, while the background was
extracted with annuli centered on the source position with radii of
4\farcs4 and 40$^{\prime\prime}$. The source to the east of \pmn (no.
1, see Fig.~\ref{fig-chandra}) was extracted with an extraction radius of $4\farcs9$, while the
background annulus radii were $8\farcs3$ and $25\farcs9$. 
The source to the east is seen in archival \XMM data as well. The
source to the west is not detected in \XMM, either due to its low flux,
or it is variable. 
The spectral analysis was performed with the Interactive Spectral
Interpretation System \citep[ISIS, version 1.6.2-40,][]{Houck2000}.
The \Chandra data were modeled with an absorbed powerlaw, which fits
both \pmn and the eastern source (no. 1).
Due to low S/N in the individual bins, we use Cash statistics
\citep{Cash1979} for spectral analysis.
For the absorbing column we use the abundances of
\citet{Wilms2000} and the cross sections of \citet{Verner1996} with
the newest version of the tbnew model\footnote{available online at:
  \url{http://pulsar.sternwarte.uni-erlangen.de/wilms/research/tbabs/}}.
In order to determine the coordinates of the X-ray sources we ran the CIAO
tools \texttt{mkpsfmap} (at 2 keV with 50\% enclosed counts) and
\texttt{wavdetect} with the default scales.
For the de-reddening of the optical/UV/IR data, the best-fit absorbing column
$N_\mathrm{H}$ has been converted to $A_\mathrm{V}$ from X-ray dust
scattering halo measurements of \citet{Predehl1995}, with the update by
\citet{Nowak2012} for the revised abundance of the interstellar medium
\citep[see][for a detailed explanation of the treatment of
multiwavelength data]{sedcat}. 

\subsection{\Fermi/LAT $\gamma$-ray data analysis}

For the analysis of the \Fermi/LAT $\gamma$-ray data, we used the
\Fermi Science Tools (v11r0p0) with the reprocessed Pass~8 data and the
P8R2\_SOURCE\_V6 instrument response functions.
The localization of the source is tricky with \texttt{gtfindsrc} due to
its position near the Galactic plane.
The localization we performed is similar to that in the 3FGL catalog
\citep{FGL3}, and only data above 3.2\,GeV have been used as the data
have been binned 4/decade.
The log-likelihood surface is assumed to be parabolic.
A selection of eight points is sampled in a circle around the estimated
position, which is used to estimate the five parameters for the 
ellipse. The center is then moved to the estimated maximum, a new
circle chosen at the 2 $\sigma$ radius, and the procedure is iterated
until convergence. The deviation of the fit from the measured
values defines a goodness-of-fit quantity. The curvature of the
surface is used to determine the covariance matrix, which in turn
determines the positional uncertainty ellipse. We quote the values
corresponding to a 95\% containment. The positions and uncertainties
are then used to determine the need for systematic adjustment by
comparing with a set of AGNs, which have very accurate radio positions.
To account for them we multiply the 95\% uncertainties by a factor of
1.05, and add 0.433$^\prime$ in quadrature.
For the spectral analysis we used an unbinned likelihood analysis in a
region of interest of  $5^{\circ}$ around \pmn in the
1--300\,GeV energy range.
Sources within a 15$^\circ$ radius of \pmn were included in the
likelihood fitting, with their parameters fixed. A free
spectral index is used, together with a detection threshold of test
statistic $TS = 25$ \citep{Wilks1938}.

\section{Results}\label{sec-result}

\begin{figure*}\centering
  \includegraphics[width=0.88\textwidth]{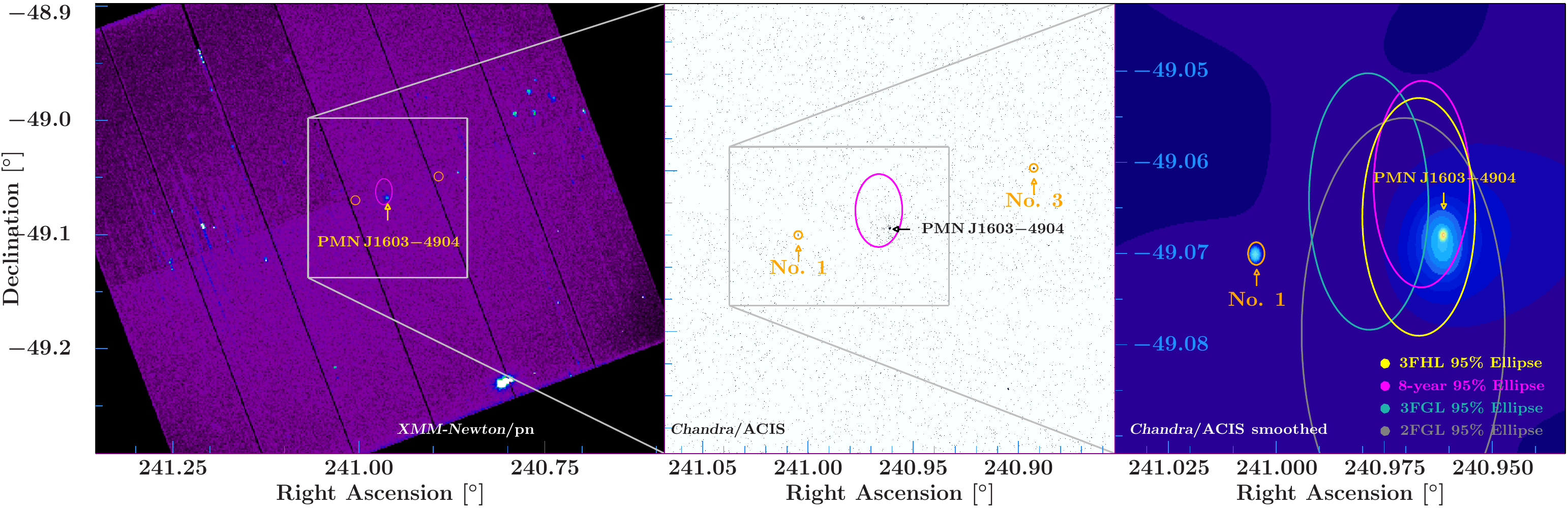}
  \caption{\XMMNewton and \Chandra/ACIS images of \pmn. The positions
    of the sources are marked with arrows. Two unknown, weak X-ray
    sources are marked in orange circles. The 95\%
    uncertainty on the \Fermi/LAT positions from the 2FGL, 3FGL, and
    the 3FHL catalogs are marked in gray, green, and yellow,
    respectively. \textsl{Left:} \XMMNewton/pn observation
    (ObsID 0724700101); \textsl{Middle:}\Chandra/ACIS observation
    (ObsID 17106); \textsl{Right:} Same \Chandra/ACIS observation, the
    image was smoothed with a Gaussian kernel of S/N range of 3--5.}
  \label{fig-chandra}
\end{figure*}

We find three X-ray sources in the \Chandra/ACIS image in the direct
vicinity of the 2MASS coordinates of \pmn \citep[see
Fig.~\ref{fig-chandra};][]{mass2}. The coordinates are given in Table~\ref{tab-pos}.
\def\arraystretch{1.35}
\begin{table}\centering\footnotesize
  \caption{\Chandra positions of the three X-ray sources in the direct vicinity
  of \pmn ( $\alpha_\mathrm{J2000.0}$ =
  $16^{\mathrm{h}}03^{\mathrm{m}}50\fs69$,
  $\delta_\mathrm{J2000.0}$ = $-49^\circ04^\prime05\farcs49 $). Positions and uncertainties have been determined using
  wavdetect. The position of the central source (no. 2) is
  consistent with the radio position of \pmn.}
  \label{tab-pos}
  \begin{tabular}{lllll}
    No. & $\alpha_\mathrm{J2000.0}$ &  $\delta_\mathrm{J2000.0}$  &
      $u(\alpha_\mathrm{J2000.0})$ & $u(\delta_\mathrm{J2000.0})$\\
    \hline
    1 & $16^{\mathrm{h}}04^{\mathrm{m}}01\fs111$ &
    $-49^\circ04^\prime12\farcs00$ & 0.139$^{\prime\prime}$ & 0.182
    $^{\prime\prime}$\\ 
    2 & $16^{\mathrm{h}}03^{\mathrm{m}}50\fs687$ &
    $-49^{\circ}04^\prime04\farcs44$ & 0.046$^{\prime\prime}$ &
    0.035$^{\prime\prime}$\\
    3 & $16^{\mathrm{h}}03^{\mathrm{m}}34\fs280$ &
    $-49^\circ02^\prime57\farcs16$ &0.238$^{\prime\prime}$ &
    0.130$^{\prime\prime}$\\     
  \end{tabular}
\end{table}
The X-ray source in the center (no. 2) matches the radio coordinates. We can exclude
the western source (no. 3) as a counterpart. It is at an angular distance of
$2\farcm9$ to \pmn, well outside the uncertainty ellipses of our analysis
and those of the \Fermi/LAT catalogs.
The eastern source (no. 1) is also outside the LAT uncertainty ellipses, but
closer, being at an angular distance of $1\farcm7$ and the 3FGL 
uncertainty region seems to be closer to the eastern source than
any of the LAT results (see Fig.~\ref{fig-chandra}, right panel). 
Based on this image, the $\gamma$-ray source is indeed the counterpart
to the young radio source, although the radio source lies just outside
the 3FGL 95\% uncertainty ellipse.
 In the 8-year \Fermi/LAT analysis, the source is detected at
 $TS=2373$, with a photon index $\Gamma=1.98\pm0.03$ and a flux
$F_{1-300\,\mathrm{GeV}}=(5.57\pm 0.22)\times
 10^{-9}$\,ph\,s$^{-1}$\,cm$^{-2}$. This flux is slightly lower than
 the fluxes reported in the LAT catalogs.

We further analyze the \Chandra spectra of \pmn and of the source to
the east of \pmn.
\begin{figure}[!h]
  \includegraphics[width=0.4\textwidth]{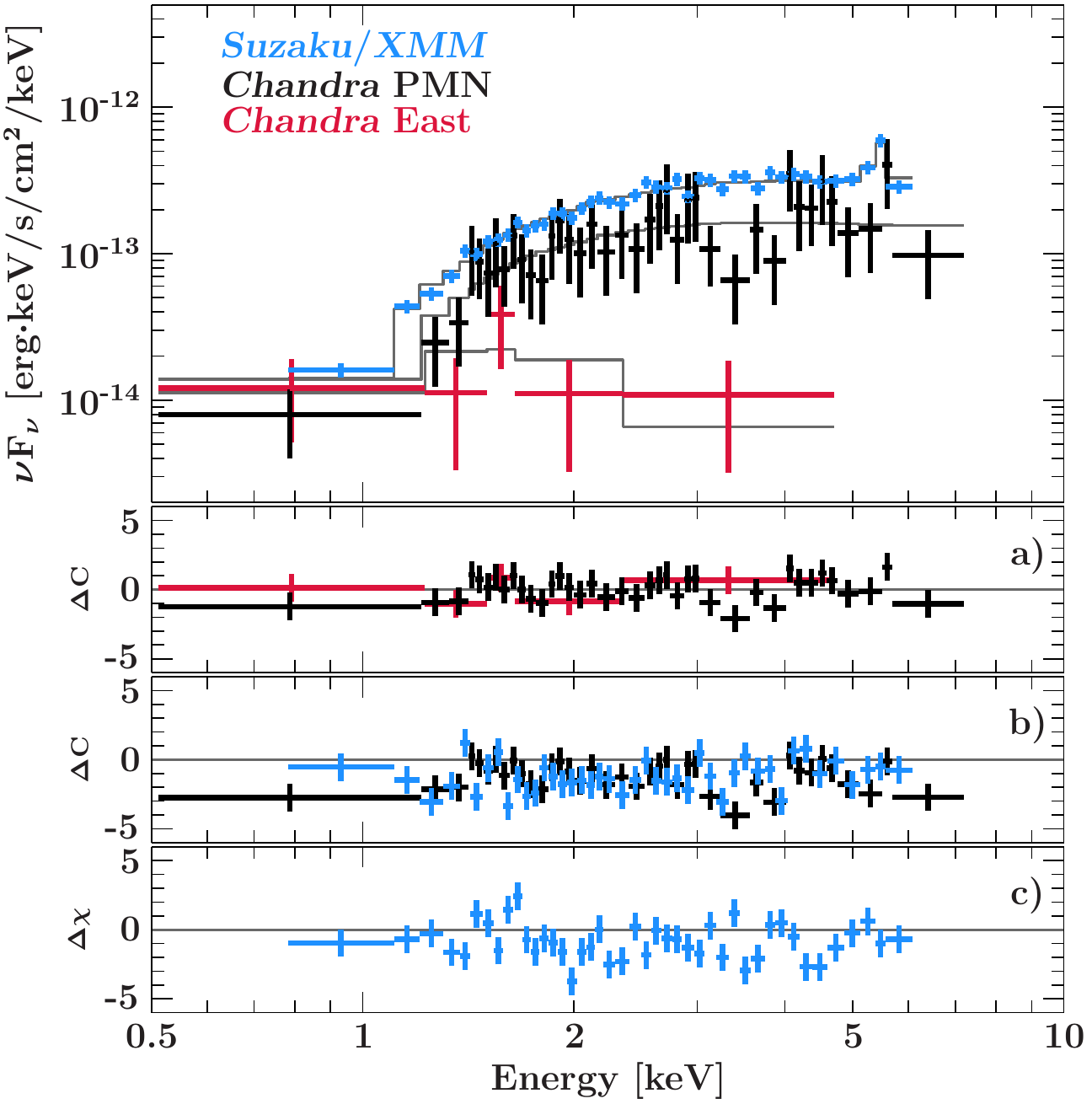}
 \caption{X-ray spectra: Combined \Suzaku/\XMMNewton
   spectrum of \pmn (blue), \Chandra/ACIS spectrum of \pmn (black),
   \Chandra/ACIS spectrum of the source east of \pmn (pink). A
   best-fit absorbed powerlaw is shown for the two \Chandra spectra.
   The fit models shown are the best fit powerlaw models for the
   \Chandra spectrum and the APEC model for the combined \Suzaku/\XMM data.
   Residuals are given for the a) best fit powerlaw spectrum,
   b) the combined APEC fit, and c) the \Suzaku/\XMM APEC fit.
 }
  \label{fig-specchandra2}
\end{figure}
X-ray observations by \Suzaku and \XMMNewton were taken from
\citet{PMN2015}. The indices are consistent with the \Chandra
best fit of \pmn, while the flux is slightly lower (see
Table~\ref{tab-fit}). Due to the low S/N, particularly at
energies above 5\,keV, the Fe emission line is not detected with
\Chandra. We model the \Chandra/ACIS data with an absorbed powerlaw and
with an absorbed collisionally-ionized emission model (APEC) and
compare the results to those of the \Suzaku/\XMMNewton data from 2013
(see Table~\ref{tab-fit2}).
\def\arraystretch{1.5}
\begin{table}\centering\footnotesize

  \caption{Best fit values from \Suzaku and \XMMNewton data taken from
  \citet{PMN2015} in comparison with best fit values for \Chandra/ACIS
  data for \object{\pmn} and two fits to the eastern source. Values without
  uncertainties have been frozen to the given value. Uncertainties are
  given at the 90\% confidence level. The absorbing column is given in
  units of $10^{22}$\,cm$^{-2}$, and the unabsorbed 2--10\,keV flux is
  given in units of $10^{-13}$\,erg\,s$^{-1}$\,cm$^{-2}$.}
  \label{tab-fit}
  \begin{tabular}{l|cccc}
    Parameter & \Suzaku \& & \Chandra & \Chandra & \Chandra\\
    & \XMM && East & East \\
    \hline
    $N_\mathrm{H}$ & $2.05^{+0.14}_{-0.12}$ & 2.05 & 2.05 & 0.632\\
    $\Gamma$ & $2.07^{+0.04}_{-0.12}$ & $2.23^{+0.29}_{-0.28}$ &
    $5.3^{+1.5}_{-2.1}$ & $3.0^{+1.4}_{-1.2}$\\
    $F_{2-10}$ & $4.39\pm0.17$ & $2.8^{+0.7}_{-0.6}$ &$0.08^{+0.05}_{-0.04}$ & $0.14^{+0.28}_{-0.1}$ \\
    
    \hline
    $\chi^2$/dof & 183.0/162 & 28.8/37 & 2.9/3  & 2.3/3\\
  \end{tabular}
\end{table}
It is worth noting that the eastern source has a very soft index of
$\Gamma=5.3^{+1.5}_{-2.1}$, assuming the same absorbing column as for
\pmn. With a 21\,cm derived Galactic equivalent hydrogen column of
$6.32\times 10^{21}$\,cm$^{-2}$ \citep{Kalberla2005} the index is
much flatter and more realistic. This suggests that the source has
little or no intrinsic absorption.
\begin{figure}[!h]
  \includegraphics[width=0.48\textwidth]{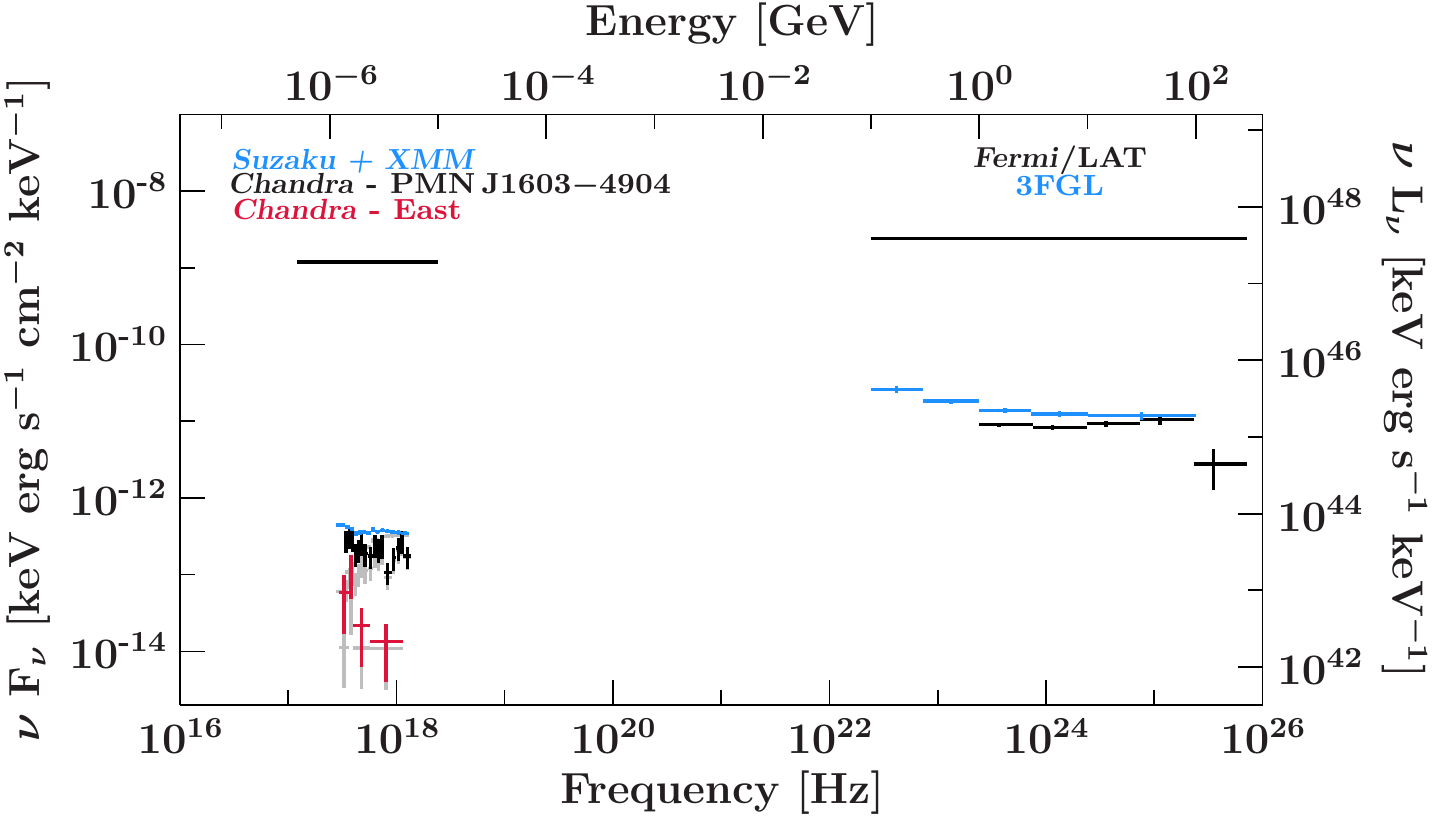}
 \caption{High-energy SED showing the archival combined \Suzaku and
   \XMM data and the \Chandra/ACIS data of both \pmn and the eastern
   source in X-rays. The \Fermi/LAT spectrum of \pmn is shown. The
   absorbed X-ray spectra are shown in gray.}
  \label{fig-sed}
\end{figure}
We add the \Chandra data to the high-energy SED, which includes the
combined \Suzaku and \XMM data, as well as the LAT spectrum from the
3FGL catalog (see Fig.~\ref{fig-sed}). We have included both the
spectrum from \pmn and from the eastern source (in black and pink,
respectively).
While it is already challenging to explain the strong 
$\gamma$-ray emission and the flat $\gamma$-ray index in combination
with the flat X-ray index of \pmn (in both \Suzaku+\XMM and \Chandra),
it is nearly impossible to explain the soft index of the eastern source in
combination with the LAT data, except by invoking different
populations of particles. The high-energy SED seems to confirm that the
eastern source is an extremely unlikely counterpart. Modeling the
broadband SED of \pmn with a physical one-zone model will remain
challenging with such a high Compton dominance and flat indices.

Although the optical/UV is likely non-thermal \citep{PMN2016op}, a
further possibility is thermal emission from an e.g., APEC component
that would explain the He-like Fe line, which is consistent with
results by \citet{Siemiginowska2016}. A combined fit to the
\Chandra/ACIS and combined Suzaku and \XMM data 
results in a best fit absorbing column of $(1.62\pm0.12)\times
10^{22}$\,cm$^{-2}$, which can explain the absence of observed He-like S/Si.
It is interesting to note that this value is smaller
than the value necessary for a purely phenomenological powerlaw fit
with an added gaussian line, which is $2.05^{+0.14}_{-0.12}\times
10^{22}$\,cm$^{-2}$. This lower value is not compatible
with the Galactic absorption of $6.32\times 10^{21}$\,cm$^{-2}$ and
suggests intrinsic absorption, possibly from a dusty torus, which is in
agreement with the interpretation of the blackbody feature in the
infrared with a hot torus \citep{PMN2014}.

\begin{table}\footnotesize
  \caption{APEC best fit values to combined fitting of \Chandra/ACIS and
    \Suzaku/\XMM data, and only \Suzaku/\XMM data. Note that the fit
    to the combined data sets was done using Cash statistic and not
    $\chi^2$ statistics, so $\chi^2$ is the Cash statistic value in
    that case. }
  \label{tab-fit2}
  \begin{tabular}{l|llll}
    Fit & $N_\mathrm{H}$ & kT  &  Abundance & $\chi^2$/dof \\
    & [$10^{22}$\,cm$^{-2}$] &[keV] & &  \\
    \hline
    comb. & $1.62\pm0.12$ & $5.9^{+0.9}_{-0.7}$ &
    $0.46^{+0.16}_{-0.14}$  & 137.298/78\\

    \XMM & $1.61\pm0.13$ & $6.1^{+1.1}_{-0.8}$ & $0.46^{+0.17}_{-0.16}$
 & 31.816/39 \\

  \end{tabular}
\end{table}
Photoionization is an alternative way to produce the He-like Fe
emission line, but the low S/N of the spectra does not allow us to
differentiate between the models.

\section{Conclusion}\label{sec-conc}
We have presented \Chandra/ACIS and \Fermi/LAT data of \pmn. We show
that we can rule out source confusion between the \Fermi/LAT source
and sources at lower energies by using the high angular
resolution of \Chandra and an improved 8-year localization of the
$\gamma$-ray source 3FGL\,J1603.9$-$4903. The positional uncertainty
is consistent with the radio coordinates and the X-ray counterpart.
An alternative X-ray counterpart at $\sim 1^\prime$ distance to \pmn can
be ruled out as a counterpart to the $\gamma$-ray source, based on the
localization and the spectral shape. Its spectral index is further
inconsistent with LAT data, as seen from the high-energy SED.
Its absorbing column is consistent with Galactic absorption (although
the low S/N makes a good fit of the absorbing column impossible).

Using \Chandra/ACIS positions, fluxes, and spectra, in combination with
\Fermi/LAT data, we confirm the X-ray counterpart, rule out
contributions from nearby X-ray sources, and confirm a high Compton
dominance in the broadband SED of \pmn.
We therefore conclude that \pmn is likely one of only two known
$\gamma$-ray emitting young radio galaxies and that its emission
mechanisms and strong Compton dominance warrant further research.


\begin{acknowledgements}
  We thank the anonymous referee for helpful comments.
  We thank F. D'Ammando as the internal LAT referee and A. Dominguez,
  E. Cavazutti, and D. Thompson for helpful comments which have greatly
  improved the paper. 
  We thank M. Hanke for the help with the projections of the TS maps.
  F. K. acknowledges funding from the European Union’s Horizon 2020
  research and innovation program under grant agreement No 653477.
  C.M. acknowledges funding from the ERC Synergy Grant ``BlackHoleCam:
  Imaging the Event Horizon of Black Holes'' (Grant 545 610058).  
  We thank J.E.~Davis for the development of the \texttt{slxfig}
  module that has been used to prepare the figures in this work.
  This research has made use of a collection of
  \ISIS scripts provided by the Dr. Karl Remeis-Observatory, Bamberg,
  Germany at \url{http://www.sternwarte.uni-erlangen.de/isis/}.
  The \textit{Fermi}-LAT Collaboration acknowledges support for LAT
  development, operation and data analysis from NASA and DOE (United
  States), CEA/Irfu and IN2P3/CNRS (France), ASI and INFN (Italy),
  MEXT, KEK, and JAXA (Japan), and the K.A.~Wallenberg Foundation, the
  Swedish Research Council and the National Space Board (Sweden).
  Science analysis support in the operations phase from INAF (Italy)
  and CNES (France) is also gratefully acknowledged. This work
  performed in part under DOE Contract DE-AC02-76SF00515.
 \end{acknowledgements}


\end{document}